\documentclass[12pt]{article}
\usepackage[margin=1in]{geometry}
\usepackage{setspace}
\usepackage{graphicx,amsmath,amsfonts}
\usepackage{setspace}
\usepackage{bbm}
\usepackage{natbib}
\usepackage{multirow}
\usepackage{hyperref}
\usepackage{amsmath}
\usepackage{amsbsy}
\usepackage{hyperref}
\usepackage{ragged2e}
\newtheorem{theorem}{Theorem}[section]

\DeclareMathOperator*{\argmin}{arg\,min}

\begin{document}
\title{Analysis of Two-Phase Studies using Generalized Method of Moments}
\author{Prosenjit Kundu \\
\textit{Department of Biostatistics, Bloomberg School of Public Health}, \\ \textit{The Johns Hopkins University, Baltimore, MD 21205, U.S.A.}, \\ email:\href{pkundu@jhu.edu}{pkundu@jhu.edu}
\and
Nilanjan Chatterjee \\
\textit{Department of Biostatistics, Bloomberg School of Public Health, }\\ \textit{Department of Oncology, School of Medicine,}\\ \textit{The Johns Hopkins University, Baltimore, MD 21205, U.S.A.}, \\ email:\href{nchatte2@jhu.edu}{nchatte2@jhu.edu}}
\maketitle
\begin{center}
\textbf{Summary}
\end{center}
Two-phase design can reduce the cost of epidemiological studies by limiting the ascertainment of expensive covariates or/and exposures to an efficiently selected subset (phase-II) of a larger (phase-I) study. Efficient analysis of the resulting dataset combining disparate information from phase-I and phase-II, however, can be complex. Most of the existing methods including semiparametric maximum-likelihood estimator, require the information in phase-I to be summarized into a fixed number of strata. In this paper, we describe a novel method for analysis of two-phase studies where information from phase-I is summarized by parameters associated with a reduced logistic regression model of the disease outcome on available covariates. We then setup estimating equations for parameters associated with the desired extended logistic regression model, based on information on the reduced model parameters from phase-I and complete data available at phase-II after accounting for non-random sampling design at phase-II. We use the generalized method of moments to solve overly identified estimating equations and develop the resulting asymptotic theory for the proposed estimator. Simulation studies show that the use of reduced parametric models, as opposed to summarizing data into strata, can lead to more efficient utilization of phase-I data. An application of the proposed method is illustrated using the US National Wilms Tumor study data.  

\section{Introduction}
Modern epidemiological studies often require collection of information on a large number of factors, including lifestyle and behavioral factors, social and environmental conditions, and biomarkers. Measuring certain factors, such as novel biomarkers or physical activity levels based on wearable devices, can be cost-prohibitive. The difficulty can be overcome by employing a two-phase sampling design where at phase-I, a relatively large number of individuals are sampled from a target population for the ascertainment of a set of inexpensive covariates. At phase-II, a small sub-sample is then judiciously selected, possibly stratified by disease status and covariate information collected at phase-I, for the ascertainment of more expensive covariates. Two-phase sampling was first introduced by Neyman in 1938 as an approach for stratification and gradually, it gained popularity in many other fields including epidemiology, econometrics and GWAS \citep{neyman_contribution_1938, manski_estimation_1977, pickles_screening_1995, schaid_two-phase_2013, thomas_two-phase_2013}. Several studies have illustrated the design and analysis of two-phase studies using the data from the National Wilms Tumor Study \citep{dangio_treatment_1989, green_comparison_1998, breslow_design_1999}. 

Existing methods for logistic regression analysis of two-phase epidemiological studies include weighted-likelihood \citep{flanders_analytic_1991} (WL) and conditional-likelihood \citep{wild_fitting_1991, hsieh_estimation_1985, breslow_logistic_1988} (CML), which essentially focus on the analysis of the phase-II data, after accounting for sampling probability through weights or offsets, respectively. Information from phase-I data in these methods can be incorporated through post-hoc estimation of sampling weights based on available covariates. A variety of methods have been proposed to analyze two-phase designs under a semi-parametric missing data framework, where no modeling assumption regarding distribution of covariates is required. Examples include methods based on estimated-likelihood \citep{pepe_nonparametric_1991, carroll_semiparametric_1991, hu_estimation_1996}, regression calibration \citep{chen_unified_2000}, pseudo-score \citep{chatterjee_pseudoscore_2003} , weighted likelihood with weights calibrated by various sample survey techniques \citep{breslow_semiparametric_2013} and semiparametric maximum likelihood (SPML) \citep{robins_estimation_1994, breslow_maximum_1997, scott_fitting_1997, lawless_semiparametric_1999, zhou_statistical_2011, qin_using_2015}.

In this article, we address two major challenges associated with the existing methods. First, a variety of the existing methods assume that the available phase-I data can be summarized into a finite number of strata and as a result, they cannot effectively utilize information available on continuous covariates at phase-I. For example, many researchers have proposed semiparametric maximum likelihood estimation, but these methods are only efficient under the assumption that the phase-I data can be summarized into finite strata \citep{breslow_large_2003}. Another challenge for the analysis of two-phase studies can arise in the setting of large consortium based studies that require data sharing. For example, large consortia have been formed for conducting GWAS of various diseases. In such consortia, studies often share individual-level data on samples (e.g. a case-control sample) which are genotyped, but individual-level data from the large underlying study (e.g. a cohort study) is not typically made available. In such a setting, it may be possible to get some summary-level information, such as estimates of parameters associated with a reduced model including some basic covariates. Thus, methods that can incorporate summary-level data from phase-I data can facilitate the incorporation of two-phase design methodology in consortia setting. 

We propose a method for the analysis of two-phase studies with a binary outcome where phase-I data can potentially involve numerous covariates, some of which could be continuous. We summarize the information from phase-I data through parameters associated with the fitting of a reduced logistic regression model. We then use the individual-level data from phase-II and estimates of the reduced model parameters from phase-I to set up a set of estimating equations for inference on parameters associated with an extended logistic regression model of interest. We use the generalized method of moment (GMM) techniques for parameter estimation and asymptotic inference. Through simulation study and real data analysis, we show that the proposed method has the same efficiency as SPMLE when the phase-I data are discrete and yet it provides more flexibility to efficiently incorporate richer phase-I data by controlling the complexity of the reduced model. 

The paper is organized as follows: in section 2, the notations and statistical formulation of the problem is described followed by asymptotic properties of the proposed estimator. In section 3, extensive simulations are conducted under different sampling designs to study the performance of the proposed method. In section 4, we illustrate applications of our method using data from the US National Wilms Tumor Study.
\section{Models and Methods}

\subsection{Model Formulation}
Let us denote the outcome of interest by $Y$, a binary variable taking values 1 and 0, and the set of full covariates by $X$, where dimension of $X$ is $q_2$. 
We assume the true relationship between $Y$ and $X$ is given by a full model (or an extended model) of the form,
\begin{equation}\label{maximal}
P(Y = i | X = x) = \frac{exp(i\beta^{T}X)}{1 + exp(\beta^{T}X)}.
\end{equation}
Our goal is to estimate and draw inference about $\beta_0$, the true value of $\beta$. Before we move onto the estimation procedure, we introduce the two-phase sampling design considered here.

\subsection{Sampling Design}
We assume at phase-I, N samples are randomly drawn from an underlying population on each of which Y and Z, a set of covariates of dimension $q_1$, are observed.  We assume $Z$ to be a subset of $X$, but it could also include surrogates of some components of $X$ where $Z$ does not have any effect on outcome Y, given X. More specifically, we will assume $Pr(Y|X,Z)= Pr(Y|X)$. Let $S :=S(Z)$ denote a set of stratifying variables in phase-I. From each of the strata defined by $Y$ and $S$ at phase-I, a random sub-sample is drawn in phase-II based on known selection probabilities denoted by $\pi(Y,S)$. 
\subsection{Method}
We first propose to summarize the phase-I data through a reduced model and use the reduced model parameters to establish an estimating equation for the full model parameters. We fit a reduced model of the form,
\begin{equation}\label{reduced_model_in_method}
Pr(Y = i|Z) = \frac{exp(\theta^TZ)}{1 + exp(\theta^TZ)}; i =0,1
\end{equation}
to the phase-I data.

We will denote $\hat{\theta}$ to be the maximum-likelihood estimator of $\theta$ and denote $\theta_0$ as the asymptotic limit of $\hat{\theta}$. Then, irrespective of whether the reduced model (\ref{reduced_model_in_method}) is correctly specified or not, we can write $E\{\mathcal{S}(Y,Z; \theta)\}|_{\theta = \theta_0} = 0$, where $\mathcal{S}(Y,Z; \theta)$ is the score function and the expectation is taken under the true data generating distribution. Assuming the maximal model (\ref{maximal}) is correct and using the law of iterated expectation by, first, taking the expectation with respect to the conditional distribution of $Y|X$, and then, marginalising over the the joint distribution of $X$, we can rewrite the score equation as $E\{f(X,Z;\beta,\theta)\}|_{\beta = \beta_0, \theta = \theta_0} = 0$, where $ f(X, Z,  \beta, \theta):= \{expit(\beta^T X) -  expit(\theta^{T}Z)\}Z$ \citep{chatterjee}. While evaluating this equation, we estimate the distribution of $X$ empirically from the individual-level phase-II data with inverse probability weighting to account for non-random sampling design. Specifically, we can rewrite the expectation by, first, taking the expectation with respect to the conditional distribution of $X|Y,S$, and then, marginalising over the the joint distribution of $(Y, S)$.  Hence, an asymptotically unbiased estimating function for $\beta$, based on summary-level data ($\hat\theta$) available from phase-I, is given by,
\begin{equation*}\label{estimatingeqnI}
 U_{1N}(\beta) = \frac{1}{N}\sum_{i = 1}^N \frac{R_i f(X_i, Z_i,  \beta, \hat{\theta})}{\pi(Y_i,S_i)},
\end{equation*}
where, $R_i$ is an indicator variable determining the selection of $i$th subject in phase-II. For rigorous derivation of the above estimating equation, see supplemental material.

Further, we propose to use the following estimating equation to incorporate data from phase-II:
\begin{equation*}
U_{2N}(\beta) = \frac{1}{N}\frac{N}{n}\sum_{i =1}^N R_i[Y_i - expit\{\gamma(\beta)^TX_i\}]X_i
\end{equation*}
The above estimating equation corresponds to standard logistic regression score equation where the effect of non-random sampling is accounted through incorporation of offset parameter in the logistic model parameter as: $\gamma(\beta) = \beta + (\log \frac{\pi(1,s)}{\pi(0,s)}, 0^T)^T$ \citep{breslow88}.

We define $\beta_{GMM}$, the GMM estimator in two-phase design,  to be the minimiser of the quadratic form, $Q_N(\beta) = U_N^T(\beta)\hat{C}U_N(\beta)$, where, $U_N(\beta) = (U^T_{1N}(\beta), U^T_{2N}(\beta))^T$ and $\hat{C}$ is a positive semi-definite matrix. Mathematically, 
$\hat{\beta}_{GMM} := \argmin_{\beta} Q_N(\beta)$. From now on, for simplicity, we denote $\hat{\beta}_{GMM} $ by $\hat{\beta}$.
 
Assume the limiting value of $\frac{n}{N}$ to be $\lambda$, where $\lambda \in (0,1)$. Let $\Psi(Y,X,R;\beta_0,\theta_0) = (\Psi_1^T, \Psi_2^T)^T$ denote the influence function of  $U_N(\hat{\boldsymbol{\beta}})$, where $\Psi_1 := \Psi_{1}(Y, X,R; \beta_0, \theta_0)  = \frac{R f(X, \beta_0, \theta_0)}{\pi(Y,S)} + \{Y - expit(\theta^T_0Z)Z\}$, $\Psi_2 :=\Psi_{2}(Y,X,R; \beta_0, \lambda) = \lambda^{-1}R\mathcal{S}(Y, X; \beta_0)$. Following the well established theory of GMM \citep{hansen, engle, imbens}, we have the following theorem:
\begin{theorem}[Consistency and Asymptotic Normality of $\hat{\beta}$]
Suppose the positive semi-definite matrix $\hat{C} \xrightarrow{P} C$. Then, under the regularity conditions (RC1-RC4) provided in appendix, $\hat{\beta} \xrightarrow{P} \beta_0$. Further, we have
\begin{center}
$\sqrt{N}(\hat{\beta} - \beta_0) \xrightarrow{D} N(0, (\Gamma^TC\Gamma)^{-1}\Gamma^TC\Delta\Omega\Delta^T C\Gamma(\Gamma^TC\Gamma)^{-1})$
\end{center}
where $\Omega = E(\Psi\Psi^T)$, $\Delta = \begin{pmatrix}
I_{q_1} & 0 & I_{q_1}\\
0 & \lambda^{-1}I_{q_2} & 0
\end{pmatrix}$ and  $\Gamma = E\frac{\partial}{\partial \beta}U(\beta, \theta)|_{\beta = \beta_0, \theta = \theta_0}$.
\end{theorem}

The above asymptotic variance is minimized at the optimal $C$ given by $C_{opt} = (\Delta \Omega \Delta^T)^{-1}$. Then, the optimal asymptotic variance is given by $(\Gamma^T(\Delta\Omega \Delta^T)^{-1}\Gamma)^{-1}$. We compute $\hat{\beta}$ using the following standard iterated GMM algorithm \citep{hansen_finite-sample_1996}. \\
\textbf{Algorithm:}  
\begin{itemize}
\item[(i)] First we choose $C$ to be an identity matrix and then minimize the quadratic form to get an initial estimate, $\hat{\beta}^{(1)}$.
\item[(ii)] Using the estimate obtained in step (i), we compute $\hat{C} =  \hat{C}_{opt} =  \{\hat{\Delta} \hat{\Omega}(\hat{\beta}^{(1)}) \hat{\Delta}^T\}^{-1}$. With this $\hat{C}$, we minimize the quadratic form to obtain $\hat{\beta}^{(2)}$.
\item[(iii)] Iterate step (ii) with the estimate obtained in step (ii) till convergence.
\end{itemize}
%
For a rigorous proof of the theorem, see supplemental material.
\section{Simulation Resembling US National Wilms Tumor Data}\label{simulation}
We conduct simulations to gain insight into the results from real data analysis (see section \ref{dataanalysis}). The data contains 4028 children diagnosed with Wilms Tumor, the most common form of kidney cancer in the pediatric age group, recruited in the third and fourth clinical trial of the National Wilms Tumor study. Details of the study can be found elsewhere \citep{dangio_treatment_1989, green_comparison_1998, breslow_design_1999}. The outcome variable of interest in this is study is relapse, a binary variable with 1 indicating that the patient's condition has deteriorated. The covariates of interest are: institutional histology (0 if favourable/1 if unfavourable); central histology (0 if favourable/1 if unfavourable); stage (0 if stage-I/1 if stage-II, 2 if stage-III and 3 if stage-IV) and age. There were two types of histology measurements available in the study. First, the institutional histology, i.e., the classification of the tumor into favorable and unfavorable, is according to the pathologist at the hospital where the children were admitted for their treatment. Because the data came from many different hospitals, it's expected that the institutional histology is likely to be more error prone due to variations associated with subjective judgements from the different pathologists. Thus, the NWTG re-evaluated histology using a central pathologist recruited for the entire study which was called central histology, the second measurement for histology available in the study. 

Imitating the structure of the real data, we assume an existence of four covariates, $X_1,X_2,X_3$, and $X_4$, where $X_1$ and $X_2$ are binary variables taking values 0 and 1; $X_3$ is an ordinal variable taking values 0,1,2 and 3; and $X_4$ is a continuous variable assumed to follow standard normal distribution. The covariates are simulated in a way such that the correlations among them are $(\rho_{12}, \rho_{13}, \rho_{14}, \rho_{23}, \rho_{24}, \rho_{34}) = (.73, .13, -.01, .09, .01, .27)$ and marginal probabilities for the discrete variables are: $Pr(X_1 = 1) = .9$, $Pr(X_2 = 1) = .89$ and $(Pr(X_3=0),Pr(X_3=1),Pr(X_3=2)) = (.39, .26, .23)$. These values are calculated from the real data. The algorithm underlying the simulation mechanism is described elsewhere \citep{amatya2015ordnor}. Let $D= (D_1,D_2,D_3)$ denote the set of dummy variables constructed for coding the variable, $X_3$, in categorical form in the underlying models. We assume the relationship between $Y$ and the covariates in the source population can be described by a logistic regression model of the form
\begin{equation*}
Pr(Y=1|X_2,D,X_4) = h(\beta_0 + \beta_1 X_2 + \beta_2^TD + \beta_3X_4 + \beta_4^TD \otimes X_2 + \beta_5X_2X_4 + \beta_6^TD \otimes X_4)
\end{equation*}
where, $ \beta_1 = 1.16$, $\beta_2 = (\beta_{2A} ,\beta_{2B} ,\beta_{2C})= (.60,.46,.81)$, $\beta_3 = .22$, $\beta_4 = (\beta_{4A} ,\beta_{4B} ,\beta_{4C} ) = (.44, 1.03, 1.63)$, $\beta_5 = -.67$, $\beta_6 = (\beta_{6A} ,\beta_{6B} ,\beta_{6C} ) = (.20, .33, .06)$ and $h(.) = (1 + (exp(.))^{-1})^{-1}$. These values are chosen by fitting the above model to the real data. The intercept parameter, $\beta_0$, is chosen to be -3.6 yielding a disease prevalence of 6\%. 

According to the above simulation scheme, we generated $10,000$ individuals in phase-I. We considered two sampling designs, a simple case-control design and a balanced design. Under the case-control design, an equal number of samples are randomly drawn from the two strata, $Y=1$ and $Y=0$, respectively. In the balanced design, we draw random samples jointly stratified by $Y$ and $X_1$ so that the resulting sample is balanced across both the levels of $Y$ and the levels of $X_1$. Previous studies \citep{breslow_logistic_1988, breslow_design_1999} have shown that a balanced sampling design can gain efficiency over standard case-control sampling for estimation of parameters associated with a covariate for which balancing is achieved.  
Here, we pretend that $X_2$ is not observed at phase-I. 

For data analysis using the proposed method, we considered the following two logistic regression models for summarizing the phase-I data.
\[ \text{M1 : } Pr(Y=1|X_1, D, X_4) = h(\gamma_0 + \gamma_1X_1 + \gamma_D^TD + \gamma_4X_4),\] 
\[  \text{M2 : } Pr(Y=1|X_1, D, X_4) = h(\gamma_0 + \gamma_1X_1 + \gamma_D^TD + \gamma_4X_4 + \gamma_5^TX_1 \otimes D + \gamma_6X_1X_4 + \gamma_7^TD \otimes X_4),\]
where $\gamma_D = (\gamma_{D1}, \gamma_{D2}, \gamma_{D3})$, $\gamma_5 = (\gamma_{5A}, \gamma_{5B}, \gamma_{5C})$ and $\gamma_7 = (\gamma_{7A}, \gamma_{7B}, \gamma_{7C})$.\\

For comparison, we also implemented a semiparametric maximum-likelihood estimator (SPMLE) where the phase-I data were summarized into discrete strata as stratum probabilities. The strata were defined by the sampling design. Although, information on $X_3$ from phase-I was not a part of the sampling design, however, we used it in post-stratification. Besides, we sought to categorize $X_4$ and incorporate it in the post-stratification. But, this led to over-stratification in $\sim$ 30\% of the simulations. Hence, we implemented the estimation procedure without including the information on $X_4$ from phase-I. The SPMLE was computed using the Chris and Wild's missreg package in R \citep{scott_calculating_2006}.
 
From the results shown in Table \ref{table:mixture},  we observe that GMM produced nearly unbiased estimates of the parameters, $\beta = (\beta_0,\beta_1,\beta_2, \beta_3, \beta_4, \beta_5, \beta_6)$  and their standard errors; and was able to maintain the coverage probabilities at the nominal level. We further observe that when the phase-I data were summarized using a more saturated model (M2), there was a very substantial gain in efficiency for the GMM estimator compared to SPMLE for covariates associated with $X_4$ and its interaction with other covariates. This highlights the desirable attribute of the GMM estimator that it can efficiently borrow information available from phase-I covariates. However, we also observed that when the phase-I data were summarized using a less saturated model (M1), the GMM estimator can lose substantial efficiency compared to the SPMLE for parameters associated with several covariates. 
\begin{table}[h]
\centering
\caption{Simulation Results (Imitating Real Data Structure)}
\footnotesize
\begin{tabular}{cccccccc}
\hline \hline
Design & Phase-I Covariates & Parameter & Bias(\%) & SD (ESD)  & CP  & RE \\ \hline \hline
\multirow{3}{*}{Case-Control} & \multirow{3}{*}{$(X_1,D,X_4)$} & $\beta_{1}$  & .029 &  .33 (.31) & .95 & .73 \\
& & $\beta_{2A}$ & .003 & .16 (.16) & .94  & 0.82  \\
& & $\beta_{2B}$ & .004 & .19 (.18) & .94 &  0.82  \\
& & $\beta_{2C}$ & 0.002 & .24 (.23) & .95 & .80 \\
& & $\beta_{3}$ & 0.003 & .12 (.12) & .95  & 1.00  \\
& & $\beta_{4A}$  & -0.015 & .45 (.43) & .94 & 0.66  \\
& & $\beta_{4B}$  & -.007 & .46 (.43) & .94 & 0.68  \\
& & $\beta_{4C}$   & .050 & .57 (.53) & .95 & 0.64  \\
& & $\beta_{5}$ & -.013 & .17 (.16) & .95 & 1.09  \\
& & $\beta_{6A}$ & -.009 & .17 (.17) & .94  & 1.00  \\
& & $\beta_{6B}$  & .003  & .18 (.17) & .94 & 1.00 \\
& & $\beta_{6C}$ & .003 & .22 (.20) & .95 & 1.04  \\ \hline
\multirow{3}{*}{Balanced} & \multirow{3}{*}{$(X_1,D,X_4)$} &$\beta_{1}$ & -.03 & .28 (.28) &.94  & 0.93 \\
& & $\beta_{2A}$& .004 & .16 (.16) & .96  & 0.83  \\
& & $\beta_{2B}$ & .007 & .18 (.18) & .95 &  0.84  \\
& & $\beta_{2C}$ & 0.03 & .22 (.22) & .94 &  .84 \\
& & $\beta_{3}$ & 1.29 & .13 (.13) & .95 & 1.08  \\
& & $\beta_{4A}$  & -.01 & .37 (.36) & .95 & 0.86  \\
& & $\beta_{4B}$  & -.01 & .37 (.36) & .95 & 0.88  \\
& & $\beta_{4C}$ & -.0008 & .40 (.39) & .95  & 0.89  \\
& & $\beta_{5}$ & -.02 & .12 (.13) & .96 & 1.21  \\
& & $\beta_{6A}$ & -.01 & .18 (.18) & .95 & 1.04  \\
& & $\beta_{6B}$  & -.001  & .18 (.18) & .96 & 1.04  \\
& & $\beta_{6C}$ & -.006 & .21 (.20) & .94 & 1.05  \\ \hline
 \multirow{3}{*}{Case-Control} & \multirow{3}{*}{$(X_1,D,X_4, X_1X_4,D\otimes X_2, D \otimes X_4)$} & $\beta_{1}$ &  0.017 & .298 (.284) & .94  & 0.89  \\
& & $\beta_{2A}$ & 0.005 & .148 (.156) & .95  & 0.96  \\
& & $\beta_{2B}$ & 0.006 & .175 (.173) & .94  & 0.93  \\
& & $\beta_{2C}$ & 0.022 & .215 (.205) & .95 & 0.97  \\
& & $\beta_{3}$ & 0.001 & .110 (.105) & .94  & 1.20 \\
& & $\beta_{4A}$ & -.007 & .388 (.375) & .93  & 0.88  \\
& & $\beta_{4B}$ & -.003 & .398 (.377) & .95 & 0.92  \\
& & $\beta_{4C}$  & -.0008 & .478 (.443) & .95  & 0.89  \\
& & $\beta_{5}$  & -.002 & .139 (.134) & .95  & 1.62  \\
& & $\beta_{6A}$ & -.006 & .145 (.139) & .95  & 1.40  \\
& & $\beta_{6B}$ & .006 & .149 (.141) & .95  & 1.45  \\
& & $\beta_{6C}$  & .007  & .176 (.162) & .94 & 1.65  \\ \hline
\multirow{3}{*}{Balanced} & \multirow{3}{*}{$(X_1,D,X_4, X_1X_4,D\otimes X_2, D \otimes X_4)$} & $\beta_1$  & -0.022 & .274 (.275) & .94  & 0.97 \\
 & & $\beta_{2A}$ & 0.011 & .159 (.156) & .94  & 0.88 \\
& & $\beta_{2B}$ & 0.029 & .176 (.178) & .96 & 0.87  \\
& & $\beta_{2C}$ & 0.052 & .205 (.213) & .95 & 0.94  \\
& & $\beta_3$ & -0.0006 & .106 (.114) & .93 & 1.43  \\
& & $\beta_{4A}$ & -0.019 & .352 (.359) & .95  & .91  \\
& & $\beta_{4B}$ & -0.016 & .349 (.355) & .96  & 0.93 \\
& & $\beta_{4C}$ & -0.033 & .378 (.389) & .94  & 0.96  \\
& & $\beta_5$ & -0.001 & .117 (.114) & .94 & 1.34  \\ 
& & $\beta_{6A}$ & -0.011 & .141 (.148) & .94 & 1.60 \\
& & $\beta_{6B}$ & -0.008 & .141 (.147) & .94 & 1.60  \\
& & $\beta_{6C}$ & -0.006 & .155 (.168) & .94  & 1.68  \\ \hline
 \multicolumn{8}{l}{%
  \begin{minipage}{16cm}~\\
    \footnotesize Biases, standard deviation (SD), estimated standard deviation (ESD), and coverage probabilities (CP) for GMM estimator. The last column shows relative efficiency (RE) with respect to SPMLE estimator.
  \end{minipage}%
}\\
\end{tabular}
\label{table:mixture}
\end{table}

\section{Application to US National Wilms Tumor Data}\label{dataanalysis}
In this section, we demonstrate an application of our methodology to simulated two-phase data constructed from the actual National Wilms Tumor study as described in section \ref{simulation}. Analogous to the study conducted earlier by Breslow and Chatterjee (1999) using this dataset, here we repeatedly simulate phase-II samples while keeping the phase-I sample to be fixed as the entire NWTS cohort \citep{breslow_design_1999}. 

Let $D$ and $S$ denote the outcome variable of interest, relapse status, and a stratum indicator variable for institutional histology, respectively. Let  $Z$ denote central histology and $W = (W_1,W_2,W_3)$ denote the set of dummy variables for stage, where $W=0$ denotes stage-I.  We assume the probability of relapse given all the covariates can be specified as,
\begin{equation}\label{realdatamaximal}
Pr(Y = 1|S, Z, W, Age) = h(\beta_0 + \beta_1Z + \beta_2^TW + \beta_3Age + \beta_4^TW \otimes Z + \beta_5 Z \ast Age + \beta_6 W \otimes Age)
\end{equation}
, where we implicitly assume that institutional histology has no information on relapse status other than central histology, given the rest of covariates. Since we have all the variables measured in the full cohort, we assume the ground truth to be the parameters associated with the model (3) fitted to the entire NWTS data. We simulate two-phase studies where we pretend that the institutional histology is available only at phase-II and we evaluate mean-squared errors of the GMM and SPMLE estimators around the ground truth.

Here, we describe the simulation of phase-II data. We classified all the subjects in phase-I into disjoint strata based on $D$ and $S$, where the strata specific counts are provided in Table \ref{table:freq}. We considered two different designs, case-control and balanced (see Table \ref{table:freq}). Here, the balanced design is defined in a similar way as described by Breslow \& Chatterjee \citep{breslow_design_1999} by sampling all the relapsed cases and all the patients with unfavorable histology. We simulated 1000 different phase-II samples based on each of the designs with the associated sampling probabilities given in Table \ref{table:freq}.

We summarized the phase-I data by fitting the following logistic regression model, 
\begin{equation*}
Pr(Y=1| Z_e, W, Age) = h(\theta_0 + \theta_1Z_e + \theta_2^TW + \theta_3^*Age + \theta_4^TW \otimes Z_e +  \theta_5 Z_e \ast Age + \theta_6 W \otimes Age)
\end{equation*}, to the phase-I data where $Z_e$ denotes institutional histology which is an error prone version of central histology, $Z$. From the simulated individual-level phase-II data and the information on parameter estimates, $(\hat{\theta}_0, \hat{\theta}_1, \hat{\theta}_2, \hat{\theta}_3, \hat{\theta}_4, \hat{\theta}_5, \theta_6)$, obtained from the fitted model in phase-I,  we estimated the regression parameters associated with model (\ref{realdatamaximal}) using our proposed methodology. Although, the variance-covariance matrix associated with the phase-I model parameters can be estimated from the phase-II sample, however, we estimated it from phase-I data as we have access to the entire dataset in this application. 

To compare the performance of the GMM estimator with the SPMLE estimator, we estimated the latter using the Chris and Wild's missreg package in R \citep{scott_calculating_2006}. In the estimation procedure, the stage variable was used in post-stratification to incorporate as much information as available from phase-I. Also, we attempted to include normalized age in discretized form for post-stratification to deploy maximum information from phase-I.  The categories were defined in an ad-hoc fashion, based on the indicator variable, $\mathbbm{1}_{Age > -0.18}$ where -0.18 is the median of normalized age. However, this led to the failure of the SPMLE procedure, in $\sim$ 80\% of the simulations, due to over-stratification leading some of the cells empty. Hence, we analyzed the data by including the stage variable in phase-I for post-stratification.

\begin{figure}[!htb]
	   \includegraphics[width=16cm]
	       {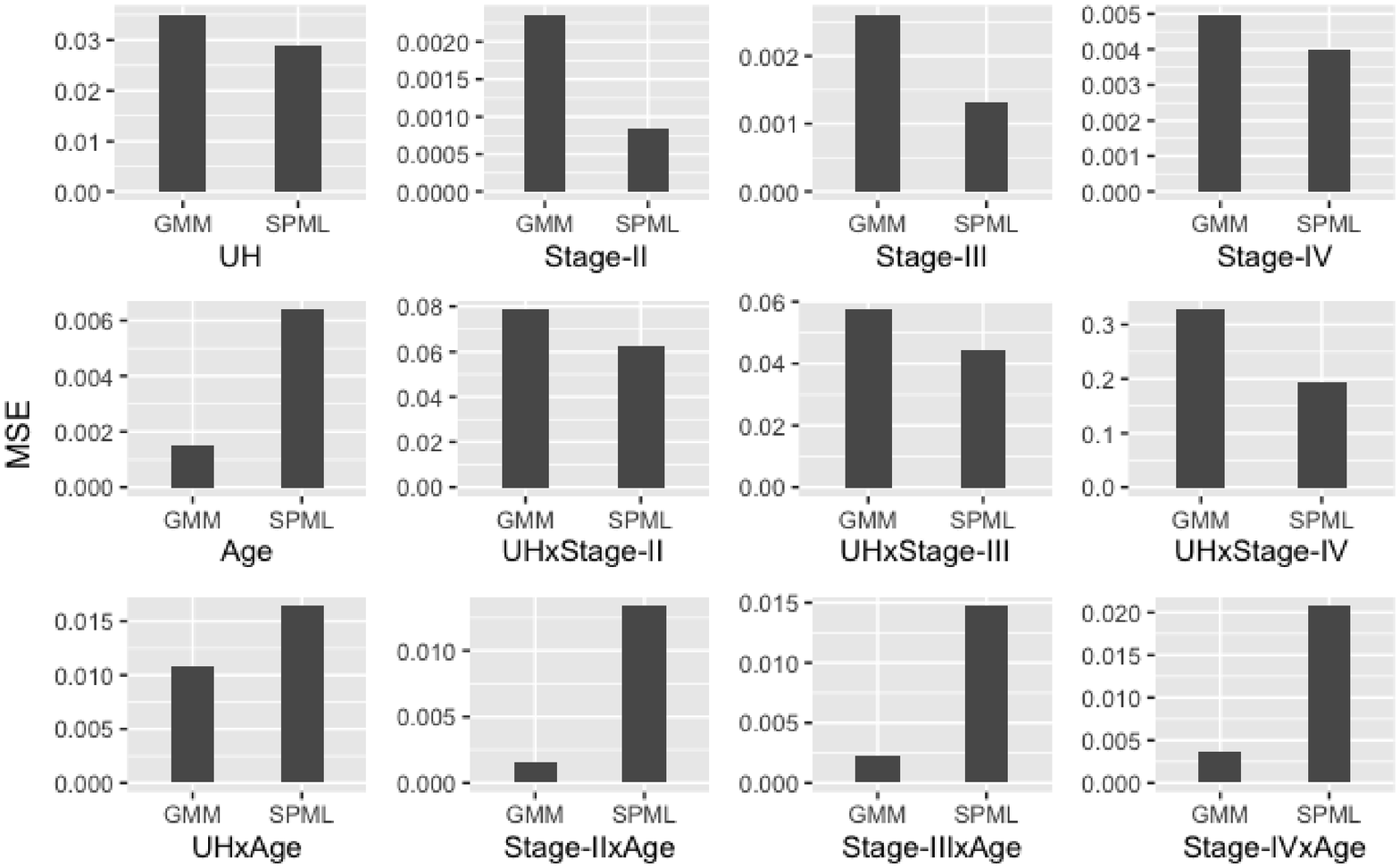}
	  \caption{Mean square errors (MSE) from real data analysis in case-control design.GMM and SPML denote the generalized method of moments and the semiparametric maximum-iklihood estimators, respectively.}
	  \label{fig:mse_cc}
\end{figure}
\begin{figure}[!htb]
	   \includegraphics[width=16cm]
	       {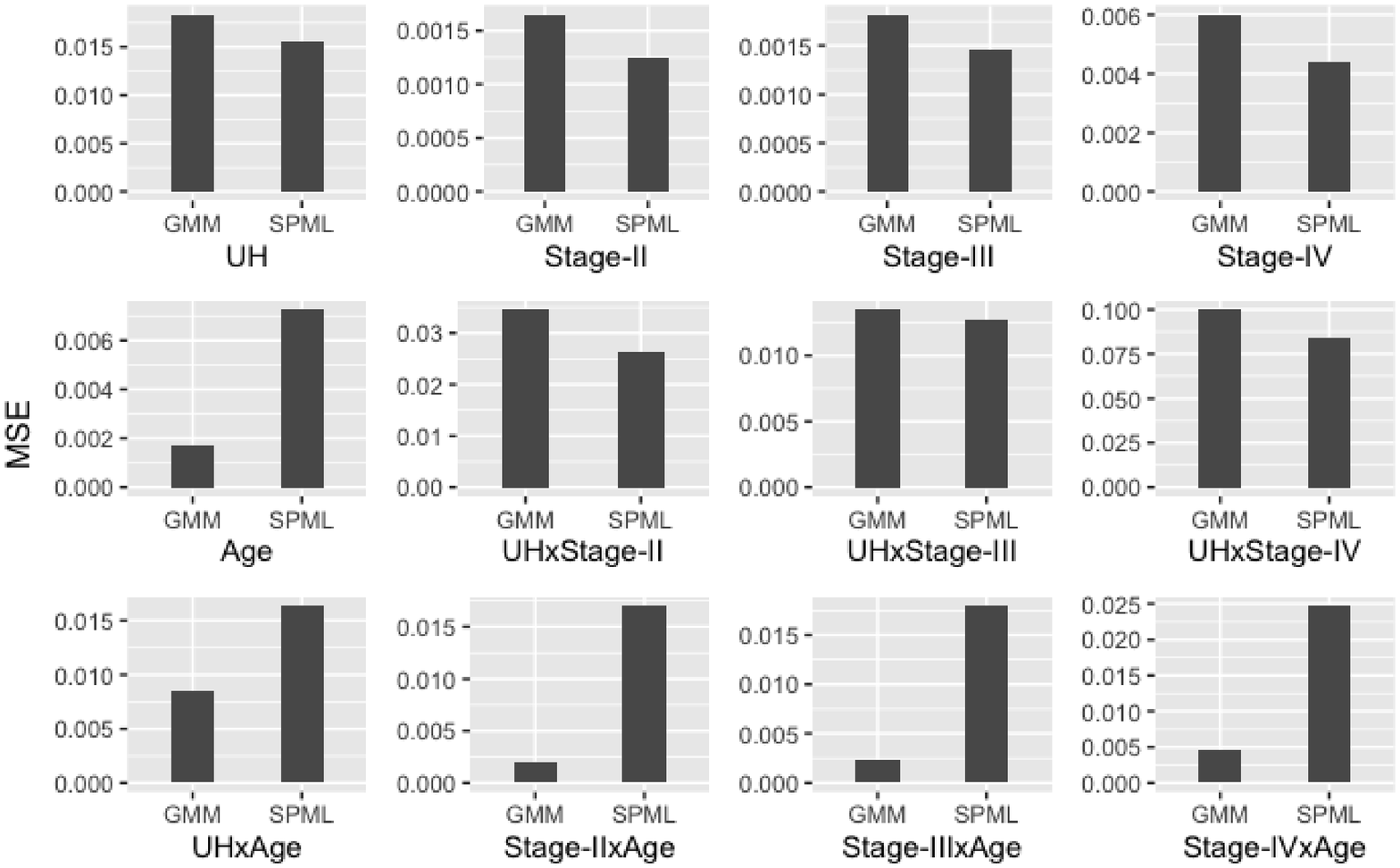}
	  \caption{Mean square errors (MSE) from real data analysis in balanced design. GMM and SPML denote the generalized method of moments and the semiparametric maximum-liklihood estimators, respectively.}
	  \label{fig:mse_bal}
\end{figure}

We calculated the mean square error of the regression coefficients around the assumed ground truth. From Figure \ref{fig:mse_cc} and \ref{fig:mse_bal}, 
we see substantial smaller MSE for the effect of age and its interaction with other covariates in both the designs. However, for some of the other covariates/terms, MSE for GMM is somewhat higher compared to SPMLE, but the loss of precision was not as much compared to the degree of gain seen for age-related coefficients. Also, we see that the loss in efficiency is alleviated for some of the covariates/terms other than age-related covariates in the balanced design compared to the case-control design.

 
\begin{table}[h]
\centering
\caption{Wilms Tumor Data: Phase-I strata frequencies and Phase-II sampling design}
\begin{tabular}{lcccccc}
\hline \hline
\multicolumn{3}{l}{\multirow{2}{*}{Phase-I: Strata Frequencies}} & \multicolumn{4}{c}{Phase-II Sampling Probabilities} \\ \cline{4-7} 
\multicolumn{3}{l}{} & \multicolumn{2}{c}{\underline{\ \ Case-Control \ \ }} & \multicolumn{2}{c}{\underline{\ \ \ \ Balanced \ \ \ \ }} \\ 
\begin{tabular}[c]{@{}l@{}}Institutional \\ Histology\end{tabular} & Cases$^a$ & Controls & Cases$^a$ & Controls & Cases$^a$ & Controls \\ \hline \hline
Favorable & 415 & 3207 & 1 & 0.165 & 1 & 0.086 \\ 
Unfavorable & 156 & 250 & 1 & 0.165 & 1 & 1 \\ \hline
 \multicolumn{7}{l}{%
  \begin{minipage}{12.4cm}~\\
    \footnotesize $^a$ Cases are defined to be the relapsed ones.
  \end{minipage}%
}\\
\end{tabular}
\label{table:freq}
\end{table}

\section{Discussion}
In this article, we have proposed a novel method for the analysis of two-phase studies which can incorporate information from complex multivariate phase-I data through summary-level parameters associated with fitted reduced models. We showed through extensive simulation studies and real data analysis that summarizing phase-I data through a set of parameters associated with an underlying reduced model, in contrast to summarizing the information into a set of strata, can lead to a more efficient way of utilizing the phase-I data in the analysis. However, the reduced model should be made as saturated as possible as the size of the data permits. Use of a highly under-specified model can result in a substantial loss of efficiency (see Table \ref{table:mixture}).

We have considered scenarios where the selection probabilities were known by design. However, in large studies with complex designs, it may be considerably difficult to retrieve true selection probabilities. One of the examples is the UKBiobank where information on physical activity and imaging are obtained on sub-cohorts but the participation of subjects in sub-cohorts is non-random and is not under the control of the investigators \citep{bycroft_uk_2018, elliott_genome-wide_2018}. Alternatively, it may be feasible to estimate the selection probabilities in a post hoc fashion, by imposing a parametric model on the selection mechanism \citep{gilbert_optimal_2014, saegusa_weighted_2013, breslow_improved_2009, little_comment:_2007, pfeffermann_role_1993}. In general, it would be a potential application of our method to such a large dataset and would be interesting to explore how the specification of such a parametric model affects the performance of GMM.


Our method relies on the generalized method of moments framework for drawing an inference. Alternatively, inference would also be conducted using empirical likelihood (EL) theory \citep{Qin2000, Qin1994} using a similar set of estimating equations. Executing the EL approach may be notably complex in spite of enjoying small sample properties. Application of EL approach in two-stage outcome-dependent sampling designs have been discussed in recent articles \citep{zhou_statistical_2011, qin_using_2015}. Computationally, the proposed method appreciates the benefits of an iterated re-weighted least squares algorithm. However, additional work is needed to compare the two frameworks.

We assumed the phase-I sample to be a random sample. However, there are many epidemiological studies that employ case-control sampling at phase-I itself \citep{breslow88, scott_fitting_1997, breslow_maximum_1997} or/and considers even more complex designs, such as multi-phase design \citep{whittemore_multi-stage_1997} and partial questionnaire designs \citep{wacholder_partial_1994} all of which creates complex missing data by design. In those scenarios, we may need to amend the estimating equations accordingly to incorporate the particular design.  Other extensions that merit future research include analysis of time-to-event outcomes based on hazard-based regression models under various two-phase sampling schemes for cohort studies, such as the case-cohort design \citep{zhou_semiparametric_2019, liu_nonparametric_2018, lin_cox_1993, wacholder_alternative_1989, prentice_case-cohort_1986}.

\section*{Acknowledgement}
Research reported in this article was supported through a Patient-Centered Outcomes Research Institute (PCORI) Award (ME-1602-34530). The statements and opinions in this article are solely the responsibility of the authors and do not necessarily represent the views of the Patient-Centered Outcomes Research Institute (PCORI),  its Board of Governors or Methodology Committee.
\section*{Software}
All the codes and simulation results are available in the Github repository through the link https://github.com/28pro92/GMMTPS.
 
\appendix
\section*{Appendix}
\emph{Regularity Conditions for Theorem 1}
\begin{itemize}
\item[(RC1):] $C$ is a positive semi-definite matrix and $E\{ CU(X; \beta, \theta_0\} = 0$ iff $\beta = \beta_0$.
\item[(RC2):] $\beta_0 \in D_{\beta}$, which is a compact.
\item[(RC3):] $E|X_iX_j| < \infty$ for all $i,j = 1,\dots, q_2-1$, where $q_2-1$ is the number of covariates in phase-II. 
\item[(RC4):] The design matrix $X$ is of full column rank. 
\end{itemize}
The last condition also ensures local identifiability of $\beta_0$ which is practically more easy to check compared to the global identification condition, (RC1).

\bibliographystyle{biom}
\bibliography{refer}

\end{document}


\title{Supplementary Material for Analysis of Two-Phase Studies using Generalized Method of Moments}
\author{Prosenjit Kundu \\
\textit{Department of Biostatistics, Bloomberg School of Public Health}, \\ \textit{The Johns Hopkins University, Baltimore, MD 21205, U.S.A.}, \\ email:\href{pkundu@jhu.edu}{pkundu@jhu.edu}
\and
Nilanjan Chatterjee \\
\textit{Department of Biostatistics, Bloomberg School of Public Health, }\\ \textit{Department of Oncology, School of Medicine,}\\ \textit{The Johns Hopkins University, Baltimore, MD 21205, U.S.A.}, \\ email:\href{nchatte2@jhu.edu}{nchatte2@jhu.edu}}
\maketitle


\section*{Proof of Theorem 1}
\textbf{Derivation of the asymptotically unbiased estimating equation from phase-I}: Applying the law of iterated expectation by, first, taking the expectation with respect to the conditional distribution of $Z|Y,S$, and then, marginalizing over the joint distribution of $(Y, S)$, we rewrite the left-hand side of the score equation as,
\begin{equation*}
\begin{split}
\hat{E}^{(I)}(S_{\hat{\theta}}(Y,Z)) 
&= \hat{E}^{(I)}_{X, S}[\hat{E}^{(I)}_{Y|X,S}\{(Y - expit(\hat{\theta}^{T}Z))Z\}] \\
&= \sum_{s=1}^J\sum_{i = 1}^n[\{(expit(\beta^Tx_i) - expit(\hat{\theta}^Tz_i^{(II)}))z_i^{(II)}\}Pr(X_i = x_i, S_i = s)]\\
&=\sum_{d=0}^1 \sum_{s=1}^J\sum_{i = 1}^n[\{(expit(\beta^Tx_i) - expit(\hat{\theta}^{T}z_i))z_i\}Pr(X_i = x_i | S_i = s, Y_i =d)\\
&\quad Pr(S_i = s, Y_i =d)]\\
&=\frac{1}{N}\sum_{i = 1}^N\sum_{d=0}^1 \sum_{s=1}^J[R_i\{(expit(\beta^Tx_i) - expit(\hat{\theta}^{T}z_i))z_i\}\frac{N_{ds}}{n_{ds}}\mathbbm{1}_{(y_i = d, s_i =s)}]\\
\end{split}
\end{equation*} where $\hat{E}$ denotes empirical expectation.
Let us denote $expit(\beta^T X_i) -  expit(\hat{\theta}^{T}Z_i)Z_i$ by $f(X_i, Z_i,  \beta, \theta)$ and $\sum_{d,s}\frac{N_{ds}}{n_{ds}}\mathbbm{1}_{\{Y_i=d, S_i=s\}}$ by $\frac{1}{\hat\pi(Y_i,S_i)}$, where $\hat\pi(Y_i,S_i) := \sum_{d,s}\frac{n_{ds}}{N_{ds}} \mathbbm{1}_{\{Y_i=d, S_i=s\}}$ denote empirical selection probabilities. Under regularity conditions, (RC2) and (RC3) in appendix and using the fact that $\hat{\pi}(d,s) = \pi(d,s) + o_p(1)$, we can replace the empirical selection probabilities with the known selection probabilities by design. Next, we show that the proposed two-stage GMM estimator is consistent and asymptotically normal similar to the proof provided in the supplementary material of our paper \citep{kundu_generalized_2019}.

\textbf{Consistency proof:} First, we show it is an asymptotically unbiased estimating equation. By weak law of large numbers (WLLN), we have $\frac{1}{N}\sum_{i = 1}^N \frac{R_i f(X_i, Z_i,  \beta, \hat{\theta})}{\pi(Y_i,S_i)} \xrightarrow{P} E\{\frac{Rf(X, Z,  \beta_0, \theta_0)}{\pi(Y_i,S_i)}\} = E_{X,Y}[E_{R|Y,X}\{\frac{Rf(X, Z,  \beta_0, \theta_0)}{\pi(Y_i,S_i)}\}]$. By MAR assumption, we have $E_{X,Y}[E_{R|Y,X}\{\frac{Rf(X, Z,  \beta_0, \theta_0)}{\pi(Y_i,S_i)}\}] = E_{X,Y}[f(X, Z,  \beta_0, \theta_0)E_{R|Y,S}\{\frac{R}{\pi(Y_i,S_i)}\}]= E_{X,Y}\{f(X, Z,  \beta_0, \theta_0)\} = 0$, under the assumption that the full or the extended model is correct.

Let $dim(\theta) = q_1$ and $dim(\beta) = q_2$. Let  $U(\beta, \theta) = (U_1^T(\beta, \theta), U_2^T(\beta))^T$, where, $U_1(\beta, \theta) = \frac{R f(X, Z;\beta, \theta)}{\pi(Y,S)}$ and $U_{2}(\beta) =  R[Y - expit\{\gamma(\beta)^TX\}]X$.
Define
$U_{0}(\beta,\theta) :=E\{U(X;\beta,\theta)\}$
and $Q_{0}(\beta) := U_{0}(\beta,\theta_0)^{T}C U_{0}(\beta,\theta_0)$. By (RC1) and Lemma 2.3 of \citet{Newey1994},
$Q_{0}(\beta)$ is uniquely minimized at $\beta_0$.
Since, $f(X, Z;\beta, \theta)$ is a continuous function in $\beta$ and $\theta$, therefore, $U_1(\beta,\theta)$ is continuous in its parameters. Also, $U_2(\beta)$ is continuous in $\beta$. In fact, we can say that $U(\beta,\theta)$ is continuous for each $(\beta,\theta) \in D_\beta \times \mathcal{N}(\theta_0)$, where $\mathcal{N}(\theta_0)$ is a compact neighbourhood around $\theta_0$. By triangle inequality, we have, $||U_1(\beta,\theta)|| \leq \frac{2R}{\pi(Y,S)} ||Z|| \leq K||Z|| $, where $K = \frac{2}{min \pi(Y,S)}$. Then, by Cauchy-Schwartz inequality and (RC3), we have $E||U_1(\beta,\theta)|| < \infty$ for each $(\beta,\theta) \in D_\beta \times \mathcal{N}(\theta_0)$. Similarly, we have $E||U_2(\beta)|| < \infty$ for each $\beta \in D_\beta$. Therefore, from lemma 2.4 of \citet{Newey1994}, $U_0(\beta,\theta)$ is continuous and $U_n(\beta,\theta)$ converges uniformly in probability to $U_0(\beta,\theta)$. Since, $\hat{\theta}$ is a consistent estimator of $\theta_0$, with probability 1, we have $\sup_{\beta \in D_\beta}||U_n(\beta,\hat{\theta}) - U_0(\beta,\hat{\theta})|| \leq \sup_{(\beta,\theta) \in D_\beta \times \mathcal{N}(\theta_0)}||U_n(\beta,\theta) - U_0(\beta,\theta)||$. Hence, from the earlier result of uniform convegence, $U_n(\beta, \hat{\theta})$ converges uniformly in probability to $U_0(\beta,\hat{\theta})$ for $\beta \in D_\beta$.

Next, we show that $U_0(\beta, \hat{\theta})$ converges uniformly in probability to $U_0(\beta, \theta_0)$. Note that, for any $\epsilon>0$, we have, with probabilty 1, $\sup_{\beta \in D_{\beta}}||U_{0}(\beta,\hat{\theta})-U_{0}(\beta,\theta_0)|| \leq \sup_{\beta \in D_{\beta}}E   (\sup_{||\theta-\theta_0||<\epsilon}||U(\beta,\theta)-U(\beta,\theta_0)||)$. By dominant convergence theorem,
$E (\sup_{||\theta-\theta_0||<\epsilon}||U(\beta,\theta)-U(\beta,\theta_0)||)$
converges to 0 for every $\beta\in D_{\beta}$ as $\epsilon$ decreases to 0.
Note that
$E (\sup_{||\theta-\theta_0||<\epsilon}||U(\beta,\theta)-U(\beta,\theta_0)||)$
decreases as $\epsilon$ decreases for each $\beta$.
By (RC2) and Dini's theorem,
$E(\sup_{||\theta-\theta_0||<\epsilon}||U(\beta,\theta)-U(\beta,\theta_0)||)$
converges uniformly in probability
to 0 for $\beta\in D_{\beta}$ as $\epsilon$ decreases to 0. Then,
$U_{0}(\beta,\hat{\theta})-U_{0}(\beta,\theta_0)$
converges uniformly in probability
to 0 for $\beta\in D_{\beta}$. Then, from the above two results, $U_n(\beta,\hat{\theta})$ converges uniformly in probabality to $U_0(\beta, \theta_0)$.

By the triangle and Cauchy-Schwartz inequalities,
\begin{align*}
    \sup_{\beta \in D_{\beta}}|Q_{n}(\beta) - Q_{0}(\beta)|
    & \leq ||\hat{C}||\sup_{\beta \in D_{\beta}}||U_{n}(\beta, \hat{\theta})-U_{0}(\beta,\theta_0)||^{2} \\
    & + 2||\hat{C}||\sup_{\beta \in D_{\beta}}||U_{0}(\beta,\theta_0)||
\sup_{\beta \in D_{\beta}}||U_{n}(\beta, \hat{\theta})-U_{0}(\beta,\theta_0)|| \\
    & + ||\hat{C}-C||\sup_{\beta\in D_{\beta}}||U_{0}(\beta,\theta_0)||^{2}
\end{align*}

Since $\hat{C}$ is a consistent estimator of $C$, hence, by continuous mapping theorem, $||\hat{C}-C||$ converges in probability to 0. Since $U_{0}(\beta,\theta_0)$ is continuous for $\beta \in D_{\beta}$
and $D_{\beta}$ is compact,
$\sup_{\beta \in D_{\beta}}||U_{0}(\beta,\theta_0)||^{2}$ is finite.
Since $\sup_{\beta \in D_{\beta}}||U_{n}(\beta, \hat{\theta})-U_{0}(\beta,\theta_0)||$ converges in probability to 0,
$\sup_{\beta \in D_{\beta}}||U_{n}(\beta, \hat{\theta})-U_{0}(\beta,\theta_0)||^{2}$ converges in probability to 0.
Thus, $Q_{n}(\beta)$ converges uniformly in probability to $Q_{0}(\beta)$ for $\beta \in D_\beta$. Recall that $\beta_0$ is the unique minimizer of $Q_{0}(\beta)$. By Theorem 2.1 of \citet{Newey1994},
$\hat{\beta}$ is a consistent estimator of $\beta_0$.

\textbf{Asymptotic normality  proof:} Next, we derive the asymptotic distribution of $\hat{\beta}$. Let $G_N(\beta) = \frac{\partial}{\partial \beta}U_N(\beta)$. Then, $G_N^T(\beta)CU_N(\beta)|_{\beta = \hat{\beta}} = 0$. By mean-value theorem, 
\begin{equation*}
\sqrt{N}U_N(\hat{\beta}) = \sqrt{N}U_N(\beta_0) + \sqrt{N}G_N(\bar{\beta})(\hat{\beta} - \beta_0)
\end{equation*}
where $\bar{\beta} \in (\beta_0, \hat{\beta})$. Pre-multiplying the above by $G_N^T(\hat{\beta})C$, we get
\begin{equation*}
\sqrt{N}(\hat{\beta} - \beta_0) = -M_N(\hat{\beta}, \bar{\beta}, \hat{\theta}) \sqrt{N} U_N(\beta_0, \hat{\theta}))
\end{equation*}
where $M_N = \{G_N^T(\hat{\beta}, \hat{\theta})CG_N(\bar{\beta}, \hat{\theta})\}^{-1}G_N^T(\hat{\beta}, \hat{\theta})C$.
Assuming $\hat{\theta}$ a consistent estimator for $\theta_0$ , we have, under some regularity conditions,  
$G_N^T(\hat{\beta}, \hat{\theta}) \xrightarrow{P} \Gamma^T$ and $G_n(\bar{\beta}, \hat{\theta}) \xrightarrow{P} \Gamma$ where $\Gamma = E_{V,X}\frac{\partial}{\partial \beta}U(\beta, \theta)|_{\beta = \beta_0, \theta = \theta_0}$. Then, $M_N \xrightarrow{P} (\Gamma^TC\Gamma)^{-1}\Gamma^TC$. Focussing on the second multiplicative term, by mean value theorem, we have 
\begin{equation*}
\sqrt{N}U_N(\beta_0, \hat{\theta}) = \sqrt{N}U_N(\beta_0, \theta_0) + \sqrt{N}V_N(\bar{\theta})(\hat{\theta} - \theta_0)
\end{equation*}
where $V_N(\bar{\theta}) = (V^T_{1N}(\beta_0, \bar{\theta}), 0^T)^T$, $\bar{\theta} \in (\hat{\theta}, \  \theta_0), \ V_{1N}(\beta_0, \bar{\theta}) = \frac{\partial}{\partial \theta}U_{1N}(\beta_0, \theta)|_{\theta = \bar{\theta}}$.
By WLLN, we have $V_{1N}(\beta_0, \bar{\theta}) \xrightarrow{P} V_1$ where $V_1 = V_1(\beta_0, \theta_0) = E\frac{\partial}{\partial \theta}U_1(\beta_0, \theta)|_{\theta = \theta_0}$

Therefore, the influence function representation is given by, 
\begin{equation*}
\begin{split}
\sqrt{N}U_N(\beta_0, \hat{\theta}) & = 
\begin{pmatrix}
I & 0 & V_{1N}\\
0 & I & 0
\end{pmatrix} \sqrt{N}
\begin{pmatrix}
U_{1N}(\beta_0, \theta_0) \\
U_{2N}(\beta_0) \\
\hat{\theta} - \theta_0
\end{pmatrix} + o_p(1) \\
& =  
\begin{pmatrix}
I & 0 & V_{1N}\mathcal{I}^{-1}_N(\bar{\theta})\\
0 & \frac{N}{n}I & 0
\end{pmatrix} 
\begin{pmatrix}
\frac{1}{\sqrt{N}}\sum_{i=1}^N \Psi_{1}(Y_i,S_i, X_i; \beta_0, \theta_0) \\
\frac{1}{\sqrt{N}}\sum_{i=1}^N \Psi_{2}(Y_i,S_i,, X_i; \beta_0) \\
\frac{1}{\sqrt{N}}\sum_{i=1}^N  \Psi_{3}(Y_i, X_i; \theta_0)
\end{pmatrix} + o_p(1) \\
& = \begin{pmatrix}
I & 0 & V_{1N}\mathcal{I}^{-1}_N(\bar{\theta})\\
0 & \frac{N}{n}I & 0
\end{pmatrix} 
\frac{1}{\sqrt{N}}\sum_{i=1}^N \Psi(Y_i,S_i, X_i; \beta_0, \theta_0) + o_p(1) 
\end{split}
\end{equation*}
where, 
$\Psi_{1}(Y_i,S_i,X_i; \beta_0,\theta_0)  = \frac{R_i f(X_i, \beta_0, \theta_0)}{\pi(Y_i,S_i)}$;
$\Psi_{2}(Y_i,S_i, X_i; \beta_0) = R_iS_{\beta_0}(Y_i, X_i)$;
$\Psi_{3}(Y_i, X_i; \theta_0) = \{Y_i - expit(\theta^T_0Z_i)\}Z_i$ and
$\mathcal{I}_N(\bar{\theta}) \ \text{is the information matrix}$.

By WLLN, $\mathcal{I}^{-1}_N(\bar{\theta}) \xrightarrow{P} \mathcal{I}^{-1}(\theta_0) = (E[h^\prime(\theta^TZ)ZZ^{T}]|_{\theta = \theta_0})^{-1}$, where $h(.) = (1 + (exp(.))^{-1})^{-1}$ Let $\frac{n}{N} \rightarrow \lambda \in (0,1)$. By central limit theorem, we have
\begin{center}
$\sqrt{N}U_N(\beta_0, \hat{\theta}) \xrightarrow{D} N(0, \Delta \Omega\Delta^T)$
\end{center}
where $\Omega = E\{\Psi(Y_i,S_i, X_i; \beta_0, \theta_0)\Psi^T(Y_i,S_i,X_i; \beta_0, \theta_0)\}$, $\Delta = \begin{pmatrix}
I_{q_1} & 0 & I_{q_1}\\
0 & \lambda^{-1}I_{q_2} & 0
\end{pmatrix}$.

Therefore, by Slutsky's theorem, we have
\begin{center}
$\sqrt{N}(\hat{\beta} - \beta_0) \xrightarrow{D} N(0, (\Gamma^TC\Gamma)^{-1}\Gamma^TC \Delta \Omega \Delta^T C \Gamma(\Gamma^TC\Gamma)^{-1})$
\end{center}
\bibliographystyle{biom}
\bibliography{refer}